\documentclass[amsmath,amssymb,showpacs,preprint]{revtex4}
\usepackage{graphicx}% Include figure files
\usepackage{bm}% bold math
\newcommand{\be}{\begin{eqnarray}}
\newcommand{\ee}{\end{eqnarray}}

\newcommand{\ba}{\begin{array}}
\newcommand{\ea}{\end{array}}

\begin{document}
\title{
Elementary gates for cartoon computation}

\author{Marek Czachor}
\affiliation{
Katedra Fizyki Teoretycznej i Informatyki Kwantowej\\
Politechnika Gda\'nska, 80-952 Gda\'nsk, Poland}

\begin{abstract}
The basic one-bit gates ($X$, $Y$, $Z$, Hadamard, phase, $\pi/8$) as well as the controlled \textsc{cnot} and
Toffoli gates are reformulated in the language of geometric-algebra quantum-like computation. Thus, all the quantum algorithms can be reformulated in purely geometric terms without any need of tensor products.
\end{abstract}
\maketitle

\section{Introduction}

Cartoon computation \cite{AC07} is a formalism for quantum-like computation based on geometric operations. One does not need tensor products to speak of entanglement, parallelism, superpositions, and interferences. The paper \cite{AC07} showed the basic principle on the Deutsch-Jozsa problem \cite{DJ}. An analogous construction was recently applied in
\cite{MO} to the Simon problem \cite{Simon}. Other oracle problems were mentioned in the context of geometric-algebra computation in \cite{P}. In the present paper I will not work with oracles but concentrate on elementary one-,  two-, and three-bit gates. This step is essential for both concrete applications and analysis of complexity of algorithms.

I first begin with explaining the link between geometric algebra and binary coding. The idea is essentially the same as in \cite{AC07}, but there are certain technical differences associated with two subsidiary dimensions (here a $(n+2)$-dimensional Euclidean space is used for coding $n$-bit numbers). Once we know how to code and perform simple operations on bits, we can introduce gates. I start with the basic one-bit gates and then introduce multiply controlled \textsc{not}s
\cite{NC}. Finally I show on a concrete example that the geometric product leads to the same type of ``compression" and parallelism as the tensor product framework of quantum computation. I end the paper with remarks on earlier approaches.

\section{Binary parametrization}

Take a $(n+2)$-dimensional Euclidean space with the basis $\{b_0,b_1,\dots,b_{n},b_{n+1}\}$. Geometric products of different basis vectors are called {\it blades\/}. One-blades (i.e. basis vectors) satisfy the Clifford algebra
\be
b_kb_l+b_lb_k=2\delta_{kl}.\nonumber
\ee
There are $2^{n+2}$ different blades. The basis vectors $b_0$ and $b_{n+1}$ play in our formalism a privileged role.
{\it Real\/} blades are those that do not involve $b_0$; the ones including $b_0$ are termed {\it imaginary\/}.
We shall see below that this terminology is consistent with a {\it complex structure\/} needed for implementation of the one-bit elementary quantum-like gates.

We shall often need in the formulas the blade $b_{n+1}$ so let us shorten the notation by $b_{n+1}=b$. The blades that do not involve $b_{n+1}$ will be termed the {\it combs\/}, and are parametrized by $n$-bit strings according to the following convention \cite{AC07}:
$b_1=c_{0;10\dots 0}$,..., $b_n=c_{0;0\dots 1}$, $b_0b_1=c_{1;10\dots 0}$,...,
$b_0b_n=c_{1;0\dots 01}$, $b_{1}b_2=c_{0;110\dots 0}$, ...,
$b_{1}b_2\dots b_n=c_{0;1\dots 1}$, $b_0b_{1}b_2\dots b_n=c_{1;1\dots 1}$.
The combs beginning with ``$0;$" or ``$1;$", are real and imaginary, respectively.
We supplement the combs by the (real) $0$-blade $1=c_{0;0\dots 0}$.
The zeroth bit ``$A;$'', separated by the semicolon from all the other bits $A_1\dots A_n$, is not needed for coding binary numbers but only for the complex structure. Therefore, one can skip it if one explicitly works with the complex structure map $i$ introduced below.

The operation of {\it reverse\/} is denoted by $^*$ and is defined on blades by $(b_{j_1}\dots b_{j_k})^*=
b_{j_k}\dots b_{j_1}$. Now let $a=b_0b$, $a_k=b_kb$, $1\leq k\leq n$. Then
\be
a_k^* c_{A;A_1\dots A_k\dots A_n}a_k
&=&
(-1)^{A_k}c_{A;A_1\dots A_k\dots A_n}
\nonumber\\
b_kc_{A_0;A_1\dots A_k\dots A_n}
&=&
(-1)^{\sum_{j=0}^{k-1}A_j}
c_{A_0;A_1\dots A'_k\dots A_n},
\nonumber
\ee
where the prime denotes negation of a bit, i.e. $0'=1$, $1'=0$.
Negation of a $k$-th bit can be expressed in algebraic terms
\begin{widetext}
\be
{\rm n}_kc_{A;A_1\dots A_k\dots A_n}
&=&
b_ka_{k-1}^*\dots a_{1}^*a^*c_{A;A_1\dots A_k\dots A_n}aa_1\dots a_{k-1}
=
c_{A;A_1\dots A'_k\dots A_n}.
\nonumber
\ee
\end{widetext}
The complex structure is defined by
\be
{i}c_{A;A_1\dots A_n}
&=&
(-1)^{A'}c_{A';A_1\dots A_n}
\nonumber.
\ee
This definition implies the usual formulas
\be
{i}^2c_{A;A_1\dots A_n}
&=&
-
c_{A;A_1\dots A_n},
\nonumber\\
e^{{i}\phi} c_{A;A_1\dots A_n}
&=&
(\cos\phi +{i}\sin\phi) c_{A;A_1\dots A_n}.
\nonumber
\ee
${i}$ and ${\rm n}_k$ commute if $0<k$.

One has now two options: Either work with explictly real coefficients but having the number of combs doubled (by the presence of the zeroth bit), or allow for ``complex" coefficients explicitly involving the linear map $i$, and then the zeroth bit can be skipped. I prefer the second option, where the combs are parametrized by $n$ indices
$c_{A_1\dots A_n}$, since it makes the formulas  compact and quantum-looking, and all the shown bits are used for coding binary numbers. Still, for geometric purposes it is important to bear in mind that the Clifford algebra is real.

\section{Elementary gates}

A one-bit gate, $1\leq k\leq n$, is
\be
G_k &=&
\frac{1}{2}(\alpha +\beta {\rm n}_k)(1+(-1)^{A_k})
+
\frac{1}{2}(\delta +\gamma {\rm n}_k)(1-(-1)^{A_k})
\nonumber
\ee
where $\alpha=\alpha_1+\alpha_2{i}$, $\beta=\beta_1+\beta_2{i}$, $\gamma=\gamma_1+\gamma_2{i}$,
$\delta=\delta_1+\delta_2{i}$; the numbers $\alpha_1$, \dots, $\delta_2$ are real.
The link to quantum computation is that the matrix of coefficients
$
\left(
\begin{array}{cc}
\alpha & \beta\\
\gamma & \delta
\end{array}
\right)
$
should be taken in a form corresponding to an appropriate quantum gate.

Let us check the concrete gates. The three Pauli gates are
\be
X_kc_{A_1\dots A_k\dots A_n}
&=&
{\rm n}_k c_{A_1\dots A_k\dots A_n},
\nonumber\\
Y_kc_{A_1\dots A_k\dots A_n}
&=&
-{i} {\rm n}_ka_k^*c_{A_1\dots A_k\dots A_n}a_k,
\nonumber\\
Z_kc_{A_1\dots A_k\dots A_n}
&=&
a_k^*c_{A_1\dots A_k\dots A_n}a_k.\nonumber
\ee
One verifies on components the usual properties
\be
X_kc_{A_1\dots 0_k\dots A_n}
&=&
c_{A_1\dots 1_k\dots A_n},
\nonumber\\
X_kc_{A_1\dots 1_k\dots A_n}
&=&
c_{A_1\dots 0_k\dots A_n},
\nonumber\\
Y_kc_{A_1\dots 0_k\dots A_n}
&=&
-{i} c_{A_1\dots 1_k\dots A_n},
\nonumber\\
Y_kc_{A_1\dots 1_k\dots A_n}
&=&
{i} c_{A_1\dots 0_k\dots A_n},
\nonumber\\
Z_kc_{A_1\dots 0_k\dots A_n}
&=&
c_{A_1\dots 0_k\dots A_n},
\nonumber\\
Z_kc_{A_1\dots 1_k\dots A_n}
&=&
-c_{A_1\dots 1_k\dots A_n}.\nonumber
\ee
The Hadamard gate:
\be
H_kc_{A_1\dots A_k\dots A_n}
&=&
\frac{1}{\sqrt{2}}{\rm n}_kc_{A_1\dots A_k\dots A_n}
+
\frac{1}{\sqrt{2}}a_k^*c_{A_1\dots A_k\dots A_n}a_k
\nonumber\\
&=&
\frac{1}{\sqrt{2}}
\big(X_k+Z_k\big)c_{A_1\dots A_k\dots A_n}.
\nonumber
\ee
The phase and $\pi/8$ gates:
\be
S_kc_{A_1\dots A_k\dots A_n}
&=&
\frac{1}{2}(1+{i})
c_{A_1\dots A_k\dots A_n}
+
\frac{1}{2}(1-{i})a_k^*c_{A_1\dots A_k\dots A_n}a_k
\nonumber\\
T_kc_{A_1\dots A_k\dots A_n}
&=&
\frac{1}{2}(1+e^{{i}\pi/4})
c_{A_1\dots A_k\dots A_n}
+
\frac{1}{2}(1-e^{{i}\pi/4})a_k^*c_{A_1\dots A_k\dots A_n}a_k
\nonumber
\ee
Let us check the latter two on components:
\be
S_kc_{A_1\dots 0_k\dots A_n}
&=&
c_{A_1\dots 0_k\dots A_n}
\nonumber\\
S_kc_{A_1\dots 1_k\dots A_n}
&=&
ic_{A_1\dots 1_k\dots A_n}
\nonumber\\
T_kc_{A_1\dots 0_k\dots A_n}
&=&
c_{A_1\dots 0_k\dots A_n}
\nonumber\\
T_kc_{A_1\dots 1_k\dots A_n}
&=&
e^{{i}\pi/4}c_{A_1\dots 1_k\dots A_n}
\nonumber.
\ee
A general controlled two-bit gate is
\be
G_{kl}
&=&
G'_k \frac{1}{2}(1+(-1)^{A_l})+G''_k \frac{1}{2}(1-(-1)^{A_l}),
\nonumber
\ee
where $G'_k$ and $G''_k$ are one-bit gates. Control-\textsc{not} (\textsc{cnot}) reads
\be
{\rm cn}_{kl}
&=&
\frac{1}{2}(1+(-1)^{A_l})+X_k \frac{1}{2}(1-(-1)^{A_l}).
\nonumber
\ee
This can be generalized to arbitrary numbers of controlling bits. An example is given by the three-bit
control-\textsc{cnot} (Toffoli) gate
\be
{\rm cn}_{klm}
&=&
\frac{1}{2}(1+(-1)^{A_m})+ {\rm cn}_{kl}\frac{1}{2}(1-(-1)^{A_m}).
\nonumber
\ee
\section{Geometric meaning of the gates}

The gates such as $H_k$ or ${\rm cn}_{kl}$ and ${\rm cn}_{klm}$ consist of {\it pairs\/} of operations, a fact suggesting that composition of $N$ gates will require $2^N$ operations. The problem is, however, more subtle. In order to see the subtlety we have to get used to thinking of all the geometric-algebra operations in geometric terms.

\subsection{Two bits, gates $X_1$ and $X_2$}

For two bits the geometric background is provided by a plane spanned by some orthonormal basis $\{e_1,e_2\}$. The blades are: $1=\circ$ (a ``charged" point), $e_1=\rightarrow$, $e_2=\uparrow$ (oriented line segments), $e_{12}=\Box$ (an oriented plane segment). The action of the gates is: $X_1c_{A_1A_2}=c_{A_1'A_2}$, $X_2c_{A_1A_2}=c_{A_1A_2'}$.
We can forget about the zeroth bit (leading to a third dimension) and visualize as follows
\be
X_1
\left(
\begin{array}{c}
\circ\\
\rightarrow\\
\uparrow\\
\Box
\end{array}
\right)
&=&
\left(
\begin{array}{cccc}
0 & 1 & 0 & 0\\
1 & 0 & 0 & 0\\
0 & 0 & 0 & 1\\
0 & 0 & 1 & 0
\end{array}
\right)
\left(
\begin{array}{c}
\circ\\
\rightarrow\\
\uparrow\\
\Box
\end{array}
\right).\nonumber
\ee
One recognizes in the matrix the tensor product $\bm 1\otimes\sigma_1$.
\be
X_2
\left(
\begin{array}{c}
\circ\\
\rightarrow\\
\uparrow\\
\Box
\end{array}
\right)
&=&
\left(
\begin{array}{cccc}
0 & 0 & 1 & 0\\
0 & 0 & 0 & 1\\
1 & 0 & 0 & 0\\
0 & 1 & 0 & 0
\end{array}
\right)
\left(
\begin{array}{c}
\circ\\
\rightarrow\\
\uparrow\\
\Box
\end{array}
\right).\nonumber
\ee
Now the matrix is $\sigma_1\otimes \bm 1$. Similar representation is found if one takes a multivector $V=V_0+V_1e_1+V_2e_2+V_{12}e_{12}=(V_0,V_1,V_2,V_{12})$. Then
\be
X_1 V &=& (V_1,V_0,V_{12},V_2),\nonumber\\
X_2 V &=& (V_2,V_{12},V_0,V_1).\nonumber
\ee
Let us recall that multivectors are, from a geometrical standpoint, sets containing different  shapes, so they have a clear geometric interpretation \cite{AC07}. Simultaneously, in the context of computation, they play a role of entangled states.

\subsection{Two bits, gates $Z_1$ and $Z_2$}

$Z_1c_{A_1A_2}=(-1)^{A_1}c_{A_1A_2}$, $Z_2c_{A_1A_2}=(-1)^{A_2}c_{A_1A_2}$,
\be
Z_1
\left(
\begin{array}{c}
\circ\\
\rightarrow\\
\uparrow\\
\Box
\end{array}
\right)
&=&
\left(
\begin{array}{cccc}
1 & 0 & 0 & 0\\
0 & -1 & 0 & 0\\
0 & 0 & 1 & 0\\
0 & 0 & 0 & -1
\end{array}
\right)
\left(
\begin{array}{c}
\circ\\
\rightarrow\\
\uparrow\\
\Box
\end{array}
\right),\nonumber\\
Z_2
\left(
\begin{array}{c}
\circ\\
\rightarrow\\
\uparrow\\
\Box
\end{array}
\right)
&=&
\left(
\begin{array}{cccc}
1 & 0 & 0 & 0\\
0 & 1 & 0 & 0\\
0 & 0 & -1 & 0\\
0 & 0 & 0 & -1
\end{array}
\right)
\left(
\begin{array}{c}
\circ\\
\rightarrow\\
\uparrow\\
\Box
\end{array}
\right).\nonumber
\ee
\subsection{Two bits, gates $H_1$ and $H_2$}

Since $H_k=(X_k+Z_k)/\sqrt{2}$,
\be
H_1
\left(
\begin{array}{c}
\circ\\
\rightarrow\\
\uparrow\\
\Box
\end{array}
\right)
&=&
\frac{1}{\sqrt{2}}
\left(
\begin{array}{cccc}
1 & 1 & 0 & 0\\
1 & -1 & 0 & 0\\
0 & 0 & 1 & 1\\
0 & 0 & 1 & -1
\end{array}
\right)
\left(
\begin{array}{c}
\circ\\
\rightarrow\\
\uparrow\\
\Box
\end{array}
\right),\nonumber\\
H_2
\left(
\begin{array}{c}
\circ\\
\uparrow\\
\rightarrow\\
\Box
\end{array}
\right)
&=&
\frac{1}{\sqrt{2}}
\left(
\begin{array}{cccc}
1 & 1 & 0 & 0\\
1 & -1 & 0 & 0\\
0 & 0 & 1 & 1\\
0 & 0 & 1 & -1
\end{array}
\right)
\left(
\begin{array}{c}
\circ\\
\uparrow\\
\rightarrow\\
\Box
\end{array}
\right).\nonumber
\ee
Note that $H_2$ is represented with permuted $\rightarrow$ and $\uparrow$.

Let us stress again that although formaly one can identify certain tensor products in the above matrices, the space of states does not involve abstract tensoring of qubits, but only geometric operations in Euclidean spaces.

\subsection{Two bits, gates ${\rm cn}_{12}$ and ${\rm cn}_{21}$}

Here
${\rm cn}_{12}c_{A_10}=c_{A_10}$, ${\rm cn}_{12}c_{A_11}=c_{A'_11}$,
${\rm cn}_{21}c_{0A_2}=c_{0A_2}$, ${\rm cn}_{21}c_{1A_2}=c_{1A'_2}$.

\be
{\rm cn}_{12}
\left(
\begin{array}{c}
\circ\\
\rightarrow\\
\uparrow\\
\Box
\end{array}
\right)
&=&
\left(
\begin{array}{cccc}
1 & 0 & 0 & 0\\
0 & 1 & 0 & 0\\
0 & 0 & 0 & 1\\
0 & 0 & 1 & 0
\end{array}
\right)
\left(
\begin{array}{c}
\circ\\
\rightarrow\\
\uparrow\\
\Box
\end{array}
\right)\nonumber\\
{\rm cn}_{21}
\left(
\begin{array}{c}
\circ\\
\uparrow\\
\rightarrow\\
\Box
\end{array}
\right)
&=&
\left(
\begin{array}{cccc}
1 & 0 & 0 & 0\\
0 & 1 & 0 & 0\\
0 & 0 & 0 & 1\\
0 & 0 & 1 & 0
\end{array}
\right)
\left(
\begin{array}{c}
\circ\\
\uparrow\\
\rightarrow\\
\Box
\end{array}
\right).\nonumber
\ee

\subsection{Three bits, gates ${\rm cn}_{123}$, ${\rm cn}_{312}$, and ${\rm cn}_{231}$}

Here the only nontrivial actions are
${\rm cn}_{123}c_{A_111}=c_{A_1'11}$, ${\rm cn}_{312}c_{11A_3}=c_{11A_3'}$,
${\rm cn}_{231}c_{1A_21}=c_{1A_2'1}$. The Euclidean space is 3-dimensional. The blades involve a point $1$, three edges
$b_1$, $b_2$, $b_3$, three walls $b_{12}$, $b_{23}$, $b_{13}$, and the cube $b_{123}$.
\be
{\rm cn}_{123}c_{011}
&=&
{\rm cn}_{123}b_{23}=c_{111}=b_{123},\nonumber\\
{\rm cn}_{123}c_{111}
&=&
{\rm cn}_{123}b_{123}=c_{011}=b_{23},\nonumber\\
{\rm cn}_{312}c_{110}
&=&
{\rm cn}_{312}b_{12}=c_{111}=b_{123},\nonumber\\
{\rm cn}_{312}c_{111}
&=&
{\rm cn}_{312}e_{123}=c_{110}=b_{12},\nonumber\\
{\rm cn}_{231}c_{101}
&=&
{\rm cn}_{231}b_{13}=c_{111}=b_{123},\nonumber\\
{\rm cn}_{231}c_{111}
&=&
{\rm cn}_{231}b_{123}=c_{101}=b_{13}.\nonumber
\ee
Geometrically in 3D the Toffoli gate means squashing a cube into a square (one of its walls), or the other way around --- reconstructing a cube from a wall.  Composition of two different Toffoli gates exchanges walls of the cube, eg.
${\rm cn}_{312}{\rm cn}_{123}b_{23}={\rm cn}_{312}b_{123}=b_{12}$.

\section{Example}

As an example we take the simple but impressive application of ``quantum parallelism", where applying $n$ Hadamard gates (i.e. performing $n$ algorithmic steps) one generates a superposition of $2^n$ $n$-bit numbers. In quantum computation the operation looks as follows
\be
H^{\otimes n}|0_1\dots 0_n\rangle
&=&
\frac{1}{\sqrt{2^n}}
\Big(|0_1\rangle+|1_1\rangle\Big)
\dots
\Big(|0_n\rangle+|1_n\rangle\Big)
\nonumber\\
&=&
\frac{1}{\sqrt{2^n}}\sum_{A_1\dots A_n}|A_1\dots A_n\rangle.\nonumber
\ee
Quantum speedup comes from the fact that most of the operations have not to be performed by the computer itself but are taken care of by properties of the tensor product.

So let us take a look at an analogous calculation performed in the geometric-algebra framework:
\be
H_nc_{0_1\dots 0_n}
&=&
\frac{1}{\sqrt{2}}\Big({\rm n}_nc_{0_1\dots 0_n}
+
a_n^*c_{0_1\dots 0_n}a_n\Big)
\nonumber\\
&=&
\frac{1}{\sqrt{2}}\Big(c_{0_1\dots 1_n}
+
c_{0_1\dots 0_n}\Big)
=
\frac{1+b_n}{\sqrt{2}},
\nonumber\\
H_{n-1}H_nc_{0_1\dots 0_n}
&=&
\frac{{\rm n}_{n-1}
(1+b_n)
+
a_{n-1}^*(1+b_n)a_{n-1}}{\sqrt{2^2}}
\nonumber\\
&=&
\frac{
b_{n-1}(1+b_n)
+
1+b_n}{\sqrt{2^2}}
=\frac{(1+b_{n-1})(1+b_n)}{\sqrt{2^2}}.\nonumber
\ee
Let us note that the multivector $1+b_n$ is treated by ${\rm n}_{n-1}$ as a whole. From a Clifford-algebra point of view this is simply a single multivector. It makes no sense to treat $1+b_n$ as a combination of just
{\it two\/} blades, since a change of basis will map it into a combination of another number of blades. There exists a single geometric object represented by $1+b_n$. This general observation applies to all the universal gates introduced above, and shows how to geometrically interpret the number of steps of an algorithm.

Repeating the above procedure $n$ times we obtain
\be
H_1\dots H_nc_{0_1\dots 0_n}
&=&
\frac{(1+b_{1})\dots(1+b_n)}{\sqrt{2^n}}\label{H1}\\
&=&
\frac{1}{\sqrt{2^n}}\sum_{A_1\dots A_n}c_{A_1\dots A_n}.\label{H2}
\ee
Eq. (\ref{H1}) shows that the $n$-fold Hadamard gate involves $n-1$ Clifford-algebra multiplications. Even counting the additions in the braces as operations performed by the algorithm we arrive at $2n-1$ steps needed for producing a linear combination of $2^n$ binary numbers.

It is therefore clear that the geometric-product performs the same type of ``compression" as the tensor product. Multivectors of the form (\ref{H2}) can be acted upon with further gates, and in each single step one processes the entire set of $2^n$ numbers.

\section{Remarks on earlier approaches}

Links between qubits, spinors, entangled states, and geometric algebra were, of course, noticed a long time ago, much earlier than in
\cite{AC07}. One should mention, first of all, the pioneering works of Hestenes \cite{Hestenes} on relations between geometric algebra and relativity, and spinors in particular. In the context of quantum information theory the most important earlier papers are those by Havel, Doran, and their collaborators, cf. \cite{Havel-Doran,Havel,Doran,Doran2,Havel-Doran-Furuta,Knill}.

However, it seems that the very way of coding, that is, linking bits with multivectors, was in those works much less straightforward than the convention I work with in the present paper, and which was introduced in \cite{AC07}. In my opinion the ``old" approach can be reduced to replacing two-component complex vectors by $2\times 2$ matrices whose second column is empty. Such ``spinors" are matrices and thus can be written as linear combinations of the Pauli matrices, simultaneously maintaining the essential properties of the usual spinors or qubits. The Pauli matrices, on the other hand, can be regarded as a representation of geometric algebra of two- or three-dimensional Euclidean spaces. Multiparticle systems are introduced by replacing a three-dimensional space with a configuration space and one arrives at a multiparticle geometric algebra. The tensor product is then constructed by means of bivectors (appropriate bivectors  may commute with one another).

The approach used in \cite{AC07} and in the present paper is so different from those based on multiparticle geometric algebras that it is even difficult to find similarities. Here tensor products are not employed at any stage (of course sometimes some matrices are of a tensor product form, as we have seen in the case of $X_k$, say, but this is irrelevant for the construction) and even the ``$i$" I use is different. So the approach I advocate is clearly an alterantive to the earlier works that, at least in my opinion, have a status of a standard theory reformulated in a different language.

\acknowledgments

I am indebted to Krzysztof Giaro, Marcin Paw{\l}owski, Tomasz Magulski and {\L}ukasz Or{\l}owski for discussions.

\end{document}